# Next-Generation Multi-layer Metasurface Design: Hybrid Deep Learning Models for Beyond-RGB Reconfigurable Structural Colors

Omar A. M. Abdelraouf,[a,*] Ahmed Mousa,[b] and Mohamed Ragab[b]

[a] Institute of Materials Research and Engineering, Agency for Science, Technology, and Research (A*STAR), 2 Fusionopolis Way, #08-03, Innovis, Singapore 138634, Singapore.

[b] Institute of High Performance Computing, Agency for Science, Technology, and Research (A*STAR), 1 Fusionopolis Way, #16-16, Connexis, Singapore 138632, Singapore.

## ABSTRACT

Metasurfaces are key to the development of flat optics and nanophotonic devices, offering significant advantages in creating structural colors and high-quality factor cavities. Multi-layer metasurfaces (MLMs) further amplify these benefits by enhancing light-matter interactions within individual nanopillars. However, the numerous design parameters involved make traditional simulation tools impractical and time-consuming for optimizing MLMs. This highlights the need for more efficient approaches to accelerate their design. In this work, we introduce NanoPhotoNet, an AI-driven design tool based on a hybrid deep neural network (DNN) model that combines convolutional neural networks (CNN) and Long Short-Term Memory (LSTM) networks. NanoPhotoNet enhances the design and optimization of MLMs, achieving a prediction accuracy of over 98.3% and a speed improvement of 50,000x compared to conventional methods. The tool enables MLMs to produce structural colors beyond the standard RGB region, expanding the RGB gamut area by 163%. Furthermore, we demonstrate the generation of tunable structural colors, extending the metasurface's functionality to tunable color filters. These findings present a powerful method for applying NanoPhotoNet to MLMs, enabling strong light-matter interactions in applications such as tunable nanolasers and reconfigurable beam steering

* Corresponding author. Email address: Omar_Abdelrahman@imre.a-star.edu.sg

# 1. INTRODUCTION

Nanophotonic devices, particularly metasurfaces, have revolutionized the field of optics by significantly enhancing light-matter interactions at the nanoscale.[1-4] These ultrathin films can generate vivid structural colors due to their ability to manipulate light with high precision.[5-9] Metasurfaces achieve this by employing subwavelength structures that control the phase, amplitude, absorption, and polarization of light, resulting in vibrant colors without the need for pigments.[10-14] Beyond their aesthetic applications, metasurfaces have found use in creating high-quality factor cavities for light emission devices,[15] facilitating advancements in nonlinear optics,[16-18] and enabling breakthroughs in quantum optics.[19] For instance, metasurfaces have been used to engineer nanostructured films to amplify spontaneous emission[20] and enhance light absorption in solar cells[21-28] and photoelectrochemical cells.[29] Moreover, metasurfaces achieved efficient frequency conversion in nonlinear optics[17] and created an efficient and ultrathin quantum light source with manipulation of quantum states of light.[30] However, the use of a single metasurface layer often leads to suboptimal light-matter interactions of electric and magnetic modes, restricting the ability to manipulate light effectively and limiting the overall performance of these devices.[31-33]

The emergence of MLMs has addressed some of these limitations by introducing additional degrees of freedom to control light-matter coupling and modulate optical modes within a single nanostructure.[34-36] This advancement allows for the precise management of both near and far-field interactions of incoming light with the metasurfaces. Recent studies have demonstrated the potential of MLMs in enhancing the quality factor of metasurface cavities, improving phase gradient applications, and achieving tunable responses through the integration of hybrid materials.[37, 38] For example, researchers have developed multi-layer structures that dynamically control light phases in the GHz regime, leading to advanced beam steering applications.[39]

Despite these promising developments, the optimization of MLMs is hampered by the increased complexity and numerous parameters involved.[40] Traditional simulation techniques become increasingly inefficient and time-consuming as the design search space grows exponentially with the number of design parameters. Consequently, these conventional methods impose significant computational costs, limiting the extent of design optimization and restricting achievable performance due to the narrow range of available design choices.

In contrast, recent advancements in artificial intelligence (AI) have demonstrated remarkable potential for accelerating nanophotonic design optimization.[41] AI models excel at capturing the nonlinear nature of physical problems due to their extensive computational capacity and thus accelerating the development cycle and enabling the creation of more advanced optical devices.[42-44] For example, Huang et al. utilized machine learning to study the relationship between metasurface geometry and structural color, further employing reinforcement learning to predict optimal geometric configurations. However, their investigation was limited to a single-layer geometry, resulting in a restricted color gamut.[45] Similarly, Roberts et al. applied deep neural networks (DNNs) to aluminum pillar plasmonic metasurfaces, using pillar geometries as inputs. Their work faced limitations in structural color coverage due to the narrow dimensional range and weak light-matter interaction in the single-layer metasurface.[46] Hu et al. extended the DNN approach to a two-layer metasurface consisting of aluminum and silicon nitride, allowing for variable dimensions while maintaining fixed materials and geometric structures, which still constrained the achievable color gamut.[47] Recently, Souza et al. employed a DNN for predicting structural colors in a single-layer titanium dioxide metasurface, but their model was similarly restricted by specific material and dimensional parameters, leading to limited color coverage.[48] Despite these efforts, these models were confined to fixed type of nanostructures with a specific materials and primarily targeted single or double layers.

Here, we introduce NanoPhotoNet, an AI-based approach for the design optimization of multilayer metasurfaces (MLMs). To our knowledge, this is the first instance of employing AI for multi-layer metasurface design optimization. Our proposed deep learning model, NanoPhotoNet, is specifically designed to enhance the optimization of MLM structures. To effectively model the metasurface geometry, we incorporate convolutional layers that capture spatial and local patterns in metasurface dimensions and material optical properties. While convolutional neural networks (CNNs) excel at spatial pattern recognition, they often struggle with temporal dynamics, which are crucial for predicting the metasurface performance across different wavelengths of light. To address this challenge, NanoPhotoNet integrates a long short-term memory (LSTM) network at the output stage, allowing the model to process temporal information and improve the accuracy of design performance predictions over a range of wavelengths. Our proposed approach is capable of efficiently handling the design and optimization both single-layer and multilayer metasurfaces, as well as the capacity to predict the performance of novel materials when deployed within MLM designs.

## 2. Methods

Since NanoPhotoNet was built to solve general nanophotonic structures, the proposed nanophotonics device consists of MLMs with varying numbers of layers, widths, heights, and refractive indices for each layer, as well as different surrounding media and incident angles, as shown in Fig. 1a. The target problem of using a MLM is to demonstrate beyond RGB colors and achieve ultra-fast switching of these colors. Our training data set consists of a square-shaped pillar with different numbers of layers from one to five, and several materials with different refractive index values including silicon dioxide ($SiO_2$), zinc oxide (ZnO), aluminum oxide ($Al_2O_3$), silicon nitride ($Si_3N_4$), niobium pentoxide ($Nb_2O_5$), titanium dioxide ($TiO_2$), amorphous antimony trisulfide (a-$Sb_2S_3$), amorphous silicon (a-Si), and crystalline antimony trisulfide (c-$Sb_2S_3$). The dimensions studied are the width and height of each layer, the gap between pillars, and the period of the meta-atom. We used a uniform sampling between all these parameters during the training data generation. Table 1 shows the value and range of different parameters used in the generation of training data. The optical response of each data sample was theoretically calculated using the finite difference time domain (FDTD) method. Optical response is reflection and transmission versus wavelength range from 400 nm to 800 nm. The total number of training data generated is 10,836 and divided into three parts training data which represents 70% of total data, validation data around 15%, and test data of 15%. To enhance the AI model performance and accuracy, a data normalization of training data is necessary in a [0,1] range.

**TABLE I**: Design parameters for MLMs used in the dataset.

| Design parameters | Range |
|---|---|
| Number of metasurface layers ($N_0$) | 1~5 |
| Mesh size ($m$) | 5 ~ 10 nm |
| Period ($P$) | 200 ~ 500 nm |
| Refractive index ($n$) | 1.45 ~ 4.45 |
| Height ($H$) | 20 ~ 100 nm |
| Width ($W$) | 0.3*P ~ 0.9*P |
| Wavelength ($\lambda$) | 400 ~ 800 nm |
| Incident angle ($\theta$) | Normal |

Before the model training starts, training data must be reshaped to be compatible with the NanoPhotoNet model. Fig. 1b shows three main parts, input data, model architecture, and label data (output data). Since the metasurface is symmetric in the *x-axis* and *y-axis*, we reshaped the 3D shape of the metasurface into a 2D image and used uniform pixels with a pixel size of 10 nm. The input data is a 2D image of size 50×181 pixels and close pixels store information of metasurface dimensions and layer numbers. The grayscale of a 2D image pixels stores the value of the refractive index of each layer, substrate, and surrounding medium. The output data of the NanoPhotoNet model has two channels of reflection and transmission versus wavelength with a size of 2×1000.

The NanoPhotoNet model described herein integrates convolutional neural networks (CNNs) with long short-term memory (LSTM) units to exploit both spatial and temporal patterns of refractive index versus nanostructures dimensions effectively. The CNN serves as a feature extractor by applying layers of convolutional filters that capture spatial hierarchies within different metasurface layers of MLM. Convolutional layers, followed by pooling layers, reduce the dimensionality while preserving essential information, enabling the model to learn spatial features of width and height of each metasurface layer efficiently. The LSTM network, designed for handling sequential data, then processes the extracted feature maps from the CNN, allowing the model to capture temporal dependencies of optical reflection and transmission at each wavelength step. The combination of CNN and LSTM layers thus leverages the strengths of both architectures, making this model well-suited for an efficient nanophotonics design. Table 2 summarize model's layers and its shape. The proposed model is based on the fundamental convolution operation, which can be expressed mathematically as:

$$y_{i,j,k} = \sum_{m=0}^{K-1} \sum_{n=0}^{L-1} x_{i+m,j+n} \cdot w_{m,n,k} \qquad (1)$$

where $x$ is the input, $w$ is the convolution filter, and $y$ is the feature map. After feature extraction by the CNN, the output is fed into the LSTM, where the LSTM cell computations are governed by the following equations for the cell state $c_t$ and hidden state $h_t$:

$$c_t = f_t \odot c_{t-1} + i_t \odot \tilde{c}_t \qquad (2)$$

$$h_t = o_t \odot \tanh(c_t) \qquad (3)$$

Here, $i_t$, $f_t$, and $o_t$ represent the input, forget, and output gates, respectively. These equations capture both short-term and long-term dependencies, thereby augmenting the model's ability to handle sequential change in optical properties versus wavelength effectively.

The model is implemented using PyTorch, a highly flexible deep learning framework in Python, which allows for dynamic computation graphs and offers seamless integration with other Python libraries. To streamline the training and evaluation process, PyTorch Lightning is employed for making the process more reproducible and scalable. The model is trained and tested on a workstation with following specifications NIVDIA GPU with 8 GB GPU RAM and 256 GB RAM to accommodate the computational intensity of NanoPhotoNet model.

**TABLE II**: NanoPhotoNet model architecture with total number of 2,740,816 trained parameters

| Layer type | Output shape | Parameters |
| --- | --- | --- |
| Conv2D (1 → 32) | [20, 32, 177, 46] | 832 |
| MaxPool2D (2x2) | [20, 32, 88, 23] | 0 |
| Conv2D (32 → 64) | [20, 64, 84, 19] | 51,264 |
| MaxPool2D (2x2) | [20, 64, 42, 9] | 0 |
| Conv2D (64 → 128) | [20, 128, 38, 5] | 204,928 |
| MaxPool2D (2x2) | [20, 128, 19, 2] | 0 |
| Permute | [20, 38, 128] | 0 |
| LSTM (128 → 32) | [20, 38, 32] | 49,792 |
| Flatten | [20, 1216] | 0 |
| Fully Connected (FC) | [20, 2000] | 2,434,000 |

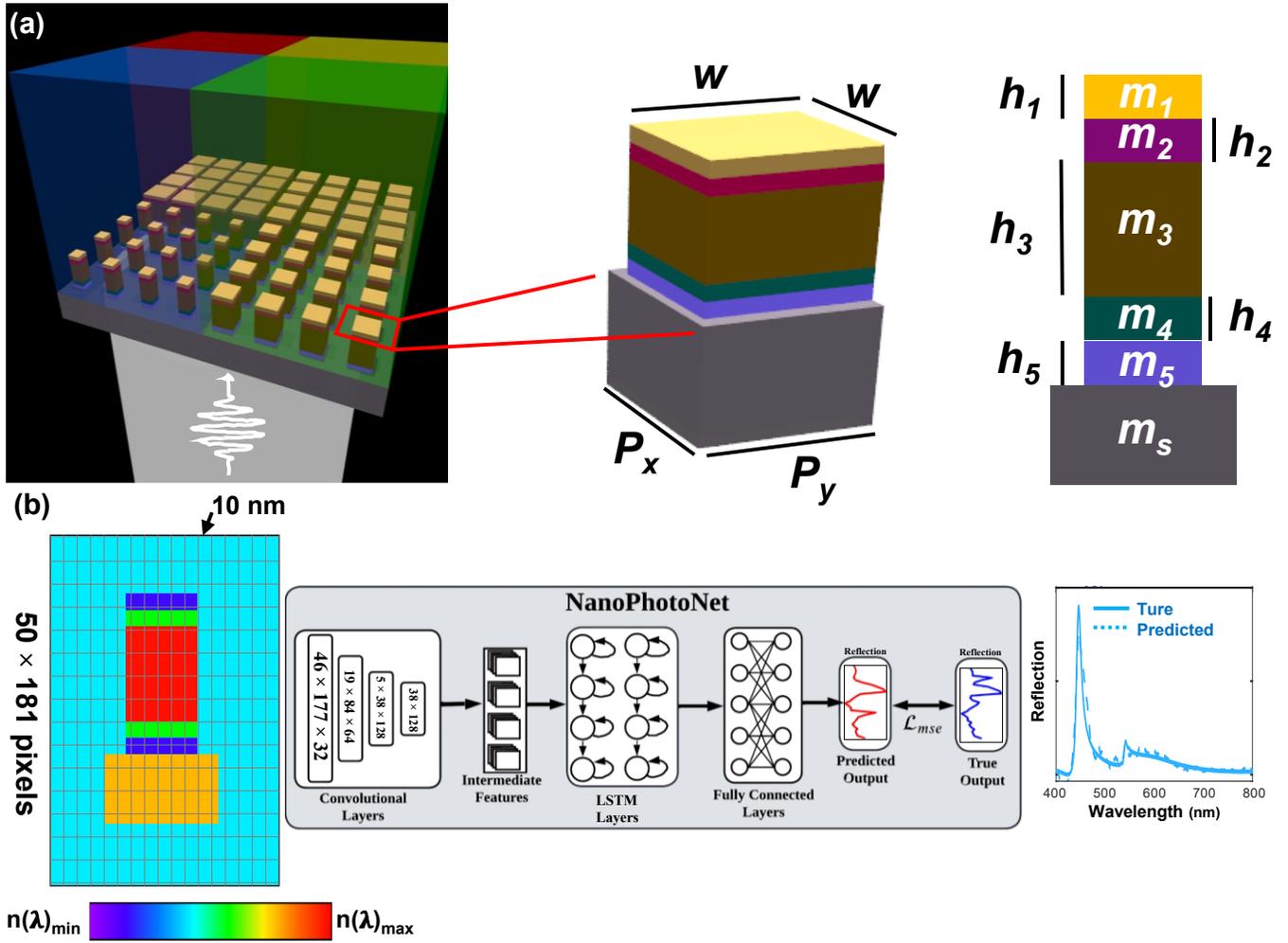

**FIG. 1.** MLMs structural color nanophotonic device. (a) Schematic of MLMs showing different emitted structural colors for different metasurface dimensions. The unit cell of MLMs has a uniform period ($P_x=P_y$) for symmetry conversion of 3D unit cell to 2D unit cell, width ($w$), and height of each layer $h_i$ where $i$ ranges from 1 to 5 and material $m_i$. (b) NanoPhotoNet model architecture with input data as a 2D image of size 50×181 pixels. The model consists of several layers of convolution, max pooling, LSTM layers, and flattened neural networks. Label data are reflection and transmission versus wavelength.

## 3. Results and Discussion

The loss of NanoPhotoNet model versus training epochs illustrated in Fig. 2. NanoPhotoNet's training and validation losses exhibit a characteristic learning curve. Initially, the training loss decreases rapidly as the model learns from the data, while the validation loss mirrors this trend but at a slightly higher value due to the model not being exposed to the validation set during training. As training progresses, the training loss tends to stabilize at a lower level, reaching a minimum value of 0.00364, while the validation loss follows a more gradual decline, with a minimum value of 0.00893. The gap between training and validation losses reflects the model's ability to generalize to unseen data.

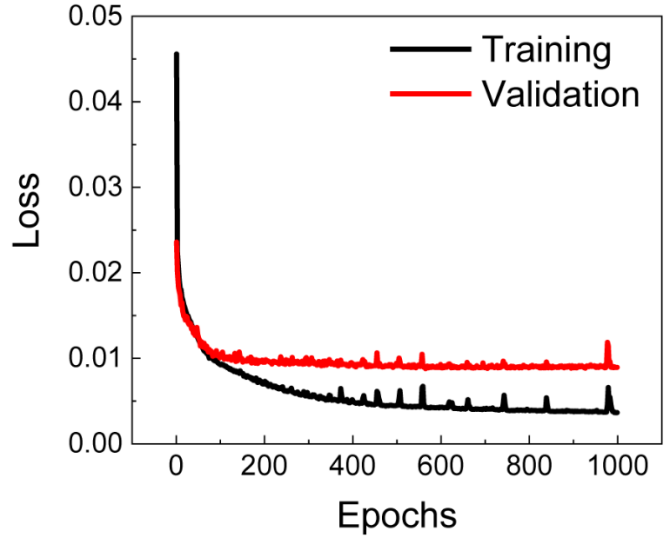

**FIG. 2.** Learning curve of NanoPhotoNet model versus training epochs.

**Section 3.1: Beyond RGB structural colors**

Beyond RGB structural color means designing a metasurface that can modulate the incident light and reflect the light with high efficiency and narrow bandwidth at certain wavelength regions. Therefore, our target is to find the best design parameters that achieve these conditions. After finishing the model training, we calculated the reflection versus wavelength to select the highest reflection and the narrowest bandwidth between all trained configurations. Using a one-layer metasurface results in a low reflection efficiency and bandwidth with color below RGB limits as observed before in the literature.[49, 50] The best configuration that achieves beyond RGB colors was using a MLM of five layers as shown in Fig. 3a. This structure has several advantages over the one-layer metasurface. First, the outer layers on top and bottom (layers number one and five) have the same refractive index to act as a Fabry-Perot cavity. Our AI model predicts using a low refractive index material ($SiO_2$) to reduce the reflection between the surrounding medium and the nanopillars. Second, the inner layers (two and four) have the same refractive index with a higher value material (ZnO) than the outer layers to reduce the refractive index mismatch between nanopillar layers. Lastly, the middle layer (three) has a very high refractive index value (a-Si) between tested materials to enhance the light-matter coupling inside the nanopillars. The optimized design parameters of the proposed MLM use thicknesses of 20 nm, 20 nm, and 100 nm for $SiO_2$, ZnO, and a-Si respectively. The simulation reflection by the FDTD model is shown in Fig. 3b. The reflection peaks are gradually tuned from a short wavelength of 400 nm to a long wavelength of 800 nm after gradually increasing the period of the proposed device from 200 nm to 400 nm width different pillars width for each color. The prediction reflection spectrums by the NanoPhotoNet model are shown in Fig. 3c. Our models showed a high prediction efficiency above 98.3% for all the simulated colors. Although, the optimum FDTD simulated data in Fig.3b. has not been seen before by the AI model. The AI calculation speed is a few milliseconds, while the FDTD simulation time is a few tens of seconds. The speed enhancement of AI model in metasurface modeling beyond structural color is more than 50,000 times faster.

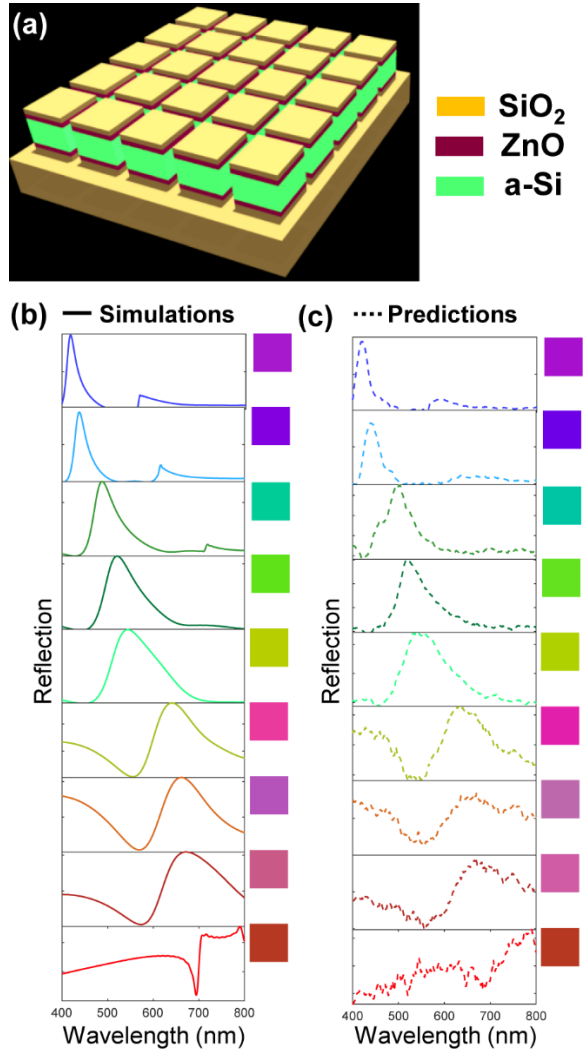

**FIG. 3.** Beyond RGB structural color using a-Si inside MLMs. (a) 3D schematic of MLMs with sandwiched a-Si layer between $SiO_2$ and ZnO layers for inducing Fabry-Perot reflection inside each pillar. (b) Simulated and predicted reflection spectrum for colors ranging from blue to red.

After comparing the reflection spectrum between FDTD calculations and NanoPhotoNet predictions and showing an acceptable prediction. We tuned several design parameters of MLMs using the same material configurations and just changed the dimension parameters to find the color gamut coverage of a-Si MLM in the CIE plot of Fig. 4a. The CIE XYZ tristimulus equations are $X = \frac{1}{k}\int R(\lambda)x(\lambda)d\lambda$, $Y = \frac{1}{k}\int R(\lambda)y(\lambda)d\lambda$, and $Z = \frac{1}{k}\int R(\lambda)z(\lambda)d\lambda$. Where $R(\lambda)$ is the reflection spectrum versus wavelength, $x(\lambda), y(\lambda),$ and $z(\lambda)$ are the color-matching functions of red, green, and blue colors respectively.[51] The maximum color coverage of a-Si MLMs is more than 137% of the RGB gamut. The CIE plot of a-Si MLM in Fig. 4a shows very pure blue, violet, green, and yellow colors outside the RGB rectangular limits. We benchmarked the NanoPhotoNet's predictions with a literature work using similar a-Si nanopillars in Fig. 4b. The color coverage of the literature work shows only a 9% RGB gamut. The color coverage of our model predicts a tremendous amount of pure color larger than the literature work with a 15 times larger color RGB gamut coverage. That indicates the strong prediction ability of the proposed NanoPhotoNet model.

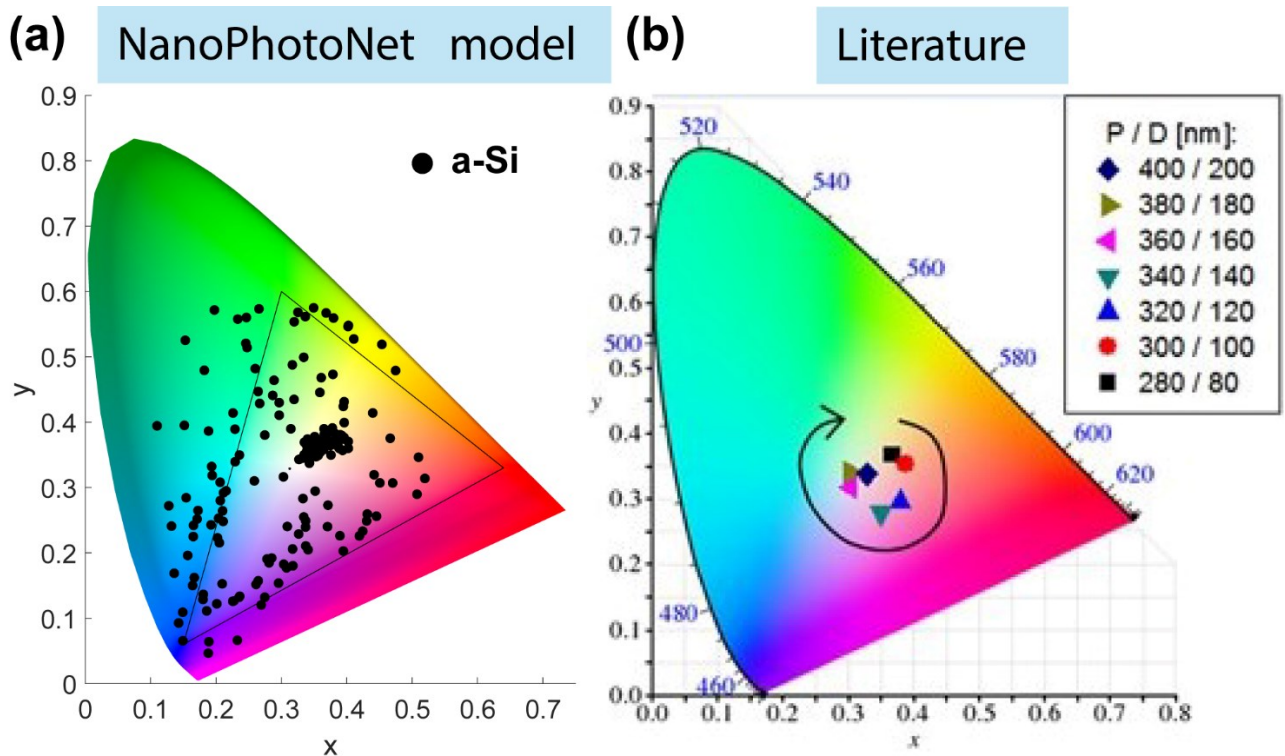

**FIG. 4.** CIE plot of a-Si MLMs designed by NanoPhotoNet model versus literature.[49] (a) a-Si MLMs with 137% color gamut beyond standard RGB colors. (b) CIE color plot of silicon nanopillars metasurfaces with different diameters and periods showing 9% RGB gamut. Reprinted with permission from Springer Nature, Copyright 2017.

**Section 3.2: Ultra-fast reconfigurable structural colors**

The second target of our study is achieving reconfigurable structural colors in MLMs with very fast switching speed. Obtaining tunable structural color requires using a material with tunable optical properties. $Sb_2S_3$ is a low-loss phase change material with a refractive index contrast between amorphous and crystalline phases of more than ($\Delta n > 1.2$). The switching between crystalline to amorphous phases of $Sb_2S_3$ is done through the melt quenching process, where a short picosecond laser pulse ps-laser randomly reorients the molecules in the crystalline lattice in Fig. 5a.[52, 53] We use the forward design approach in our trained NanoPhotoNet model to predict the possible structural colors after replacing a-Si with $Sb_2S_3$ in both phases. The FDTD simulation and NanoPhotoNet predictions using $c-Sb_2S_3$ are illustrated in Fig. 5b. The model predictions accurately show the trend of the FDTD reflection spectrum. The reflection color is gradually tuned from blue to near red by gradually scaling up the metaatom size. We observed a small broadening in the color bandwidth in the case of $c-Sb_2S_3$ because the light losses inside $c-Sb_2S_3$ are larger than a-Si, which reduces the light-matter coupling to nanopillars. After switching the phase from crystalline to amorphous, we predict the reflection spectrums of $a-Sb_2S_3$ for the same dimensions used in Fig. 5b. Fig. 5c shows the FDTD simulations and NanoPhotoNet predictions in the case of $a-Sb_2S_3$ MLM. A very good matching efficiency (>98.3%) is obtained between simulations and predictions. Using $Sb_2S_3$ improves color reflection bandwidths as the light losses of $a-Sb_2S_3$ are much lower than a-Si and $c-Sb_2S_3$. Very pure colors such as blue to green, yellow, orange, and red are obtained in $a-Sb_2S_3$ MLM.

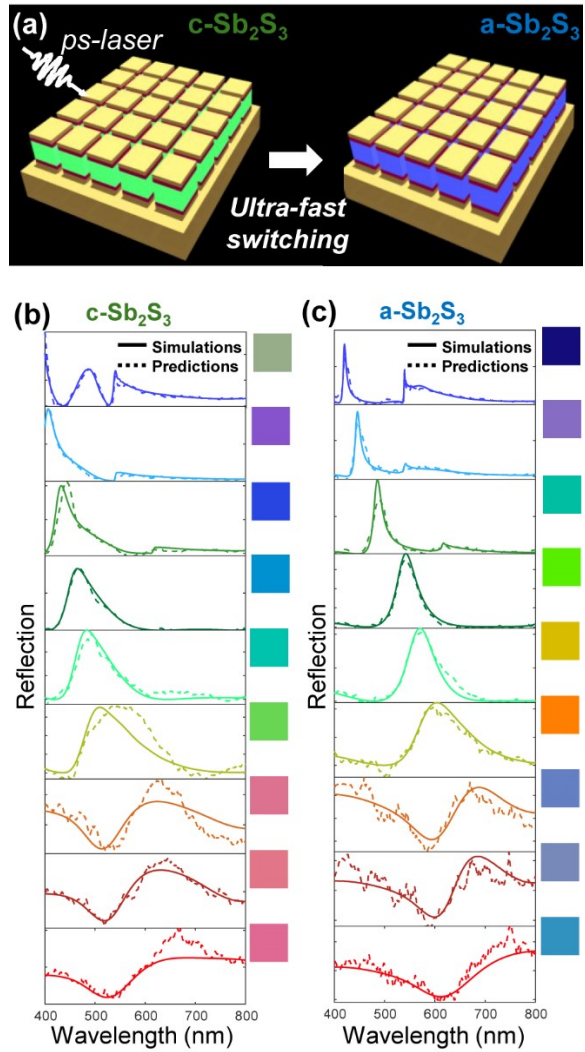

FIG. 5. Ultra-fast and reconfigurable MLMs. (a) 3D schematic of reconfigurable MLMs using $Sb_2S_3$ between $SiO_2$ and ZnO mirrors to boost light coupling inside each pillar. Ps-laser ultra-fast switches the phase of $Sb_2S_3$ from crystalline (c-$Sb_2S_3$) to amorphous (a-$Sb_2S_3$). (b) The left column shows the simulated and predicted reflected structural color for c-$Sb_2S_3$. The right column shows the simulated and predicted reflection spectrum after the phase changes to a-$Sb_2S_3$.

After obtaining ultra-fast reconfigurable color in $Sb_2S_3$ MLMs using NanoPhotoNet model, we examined different dimension values using the same layer number and materials to find the color gamut coverage in both amorphous and crystalline phases. Fig. 6a. shows the CIE plot of $Sb_2S_3$ MLMs using different phases. In the case of c-$Sb_2S_3$, the obtained color gamut is 91% RGB gamut. This reduction compared with a-Si MLM in the RGB gamut is expected as the light losses of c-$Sb_2S_3$ increase the reflection color bandwidth compared with a-Si MLM. However, the color gamut in the case of a-$Sb_2S_3$

ML increased up to 152% RGB gamut due to the reduced light losses inside a-$Sb_2S_3$ compared with a-Si and c-$Sb_2S_3$. The total obtained color gamut using both phases of $Sb_2S_3$ MLM is 163% RGB gamut. It is the highest reported color gamut in the $Sb_2S_3$-based metasurface according to our best of knowledge. Furthermore, we benchmarked our results with a literature work that uses an $Sb_2S_3$ nanopillar metasurface and changed the dimensions to obtain many colors. Fig. 6b. shows the CIE plot of the a-$Sb_2S_3$ metasurface in the literature work with only 23% RGB gamut. We believe the light-matter interactions in the reported single layer $Sb_2S_3$ metasurface were not optimized. In the case of c-$Sb_2S_3$ metasurface in the reported work, the color gamut was 22% RGB gamut in Fig. 6c. The NanoPhotoNet model predictions showed an enhancement in the color gamut in the tunable structural color by more than 7x times compared with the state-of-the-art in the literature, which shows the strong possibility of AI models to improve nanophotonic devices performance. Furthermore, we benchmarked our model accuracy with literature papers using AI models for designing structural color metasurface and listed them in Table 3. Our model shows a better accuracy and more design freedom.

**Table III:** Accuracy comparison between NanoPhotoNet model and AI models in literature.

| AI architecture | Accuracy (%) |
|---|---|
| **NanoPhotoNet** | **98.3** |
| DNN [46] | 86 |
| DNN [47] | 92.4 |

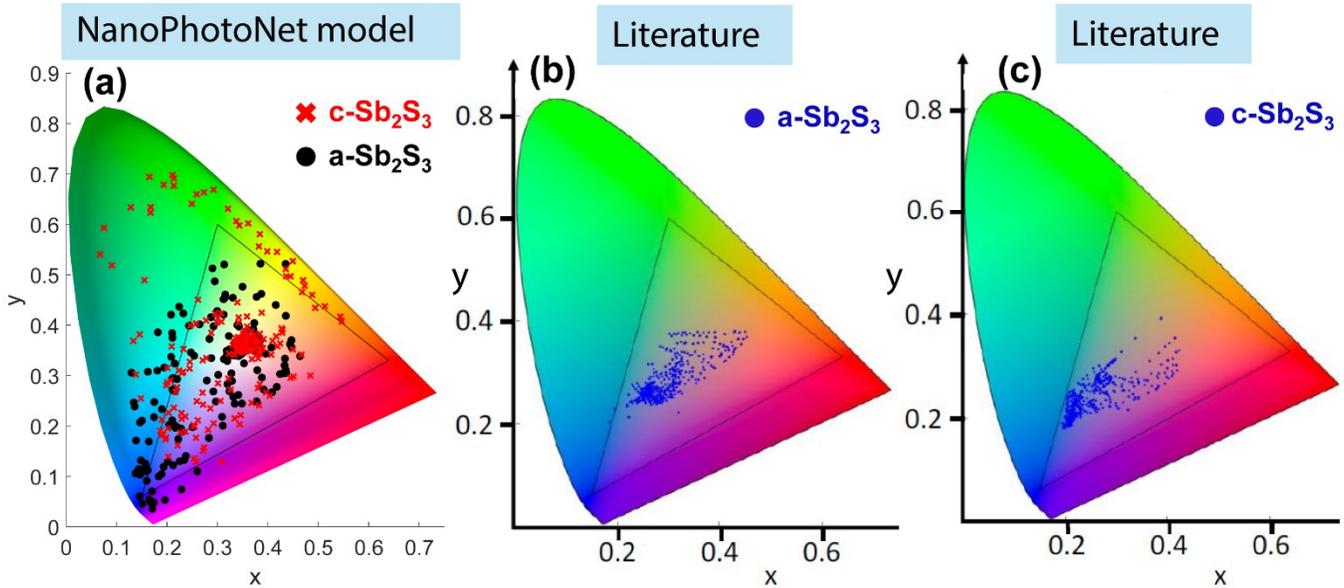

FIG. 6. CIE plot of Sb$_2$S$_3$ MLMs designed by NanoPhotoNet versus the literature.[54] (a) Predictions show an extreme light switching after phase change. The color gamut of a-Sb$_2$S$_3$ and c-Sb$_2$S$_3$ are 152% and 91% respectively. The total color gamut is up to 163% of RGB. (b) CIE plot of Sb$_2$S$_3$ nanopillars metasurface in amorphous state with 23% RGB gamut. (c) CIE plot of Sb$_2$S$_3$ nanopillars metasurface in crystalline state with 22% RGB gamut. Reprinted with permission from American Chemical Society, Copyright 2021.

## 4. Conclusion

In conclusion, we have introduced NanoPhotoNet, a hybrid deep neural network model combining CNN and LSTM architectures to optimize the design of multilayer metasurfaces. Our approach effectively addresses the limitations of traditional simulation methods by accelerating the design process and enhancing prediction accuracy for MLMs-based nanophotonic devices. NanoPhotoNet demonstrates its ability to achieve high prediction efficiency of over 98.3% while significantly improving the computational speed, resulting in a 50,000x enhancement. Furthermore, NanoPhotoNet expands the RGB gamut area by 163% and enables tunable structural colors, providing a versatile platform for advanced applications such as tunable nanolasers and reconfigurable beam steering. These results demonstrate the efficacy of NanoPhotoNet in optimizing metasurface designs and advancing the field of nanophotonics. The NanoPhotoNet model holds great potential for solving larger-scale problems in optical neural networks by enabling the design of complex, multilayered optical architectures. Its ability to efficiently handle high-dimensional design spaces and predict performance across a wide spectrum of wavelengths could significantly accelerate the development of next-generation optical computing systems.

**Credit Authorship Contribution Statement**

**Omar A. M. Abdelraouf:** Conceptualization, Supervision, Project administration, Investigation, Resources, Visualization, Validation, Resources, Methodology, Data curation, Writing – original draft & review. **Ahmed Mousa:** Data curation, Methodology, Investigation. **Mohamed Ragab:** Funding acquisition, Supervision, Project administration, Conceptualization, Methodology, Investigation, Resources, Writing – review & edition.

**Declaration of Competing Interest**

The authors declare that they have no known competing financial interests or personal relationships that could have appeared to influence the work reported in this paper.

**Data Availability**

DNN models and raw data in the paper are available on request

**Acknowledgments**

The Authors thank the Agency for Science, Technology, and Research (A*STAR) for the scholarship provided Singapore International Pre-Graduate Award (SIPGA).